\documentclass[aps,preprint,prb,showpacs]{revtex4}
\usepackage{epsfig}
\usepackage{dcolumn}
\usepackage{bm}
\usepackage{tikz}
\usepackage{graphicx, graphics, color}

\newcommand{\be}{\begin{equation}}
\newcommand{\ee}{\end{equation}}
\newcommand{\beq}{\begin{eqnarray}}
\newcommand{\eeq}{\end{eqnarray}}

\begin{document}

\title{Time reversal symmetry breaking superconductors}

\author{Karol I. Wysoki\'nski}
\email{karol@tytan.umcs.lublin.pl}
\affiliation{Institute of Physics, Maria Curie-Sk{\l}odowska University, 
20-031 Lublin, Poland}

\date{\today}

\begin{abstract}
The paper reviews  recent work on time reversal symmetry (TRS) breaking superconductors.
The family of TRS breaking superconductors is 
growing relatively fast, with many of its newly discovered members being non-centrosymmetric. 
However, many of the superconductors which possess center of inversion also break TRS. 
The TRS is often identified by means of the muon spin relaxation effect ($\mu$SR) or/and  
the Kerr effect. Both methods effectively measure the spontaneous appearance of the bulk 
magnetic field below superconducting transition temperature. One of the systems most carefully studied 
so far is Sr$_2$RuO$_4$ believed to be spin triplet chiral p-wave superconductor.
This compound provides an  example of the  material whose many 
band, multi-condensate modelling has enjoyed a number of successes. We discuss  
in some details the properties of the material. Among them is the polar Kerr effect, which understanding has 
resulted in the  discovery of the novel mechanism of the phenomenon. The mechanism is universal and 
thus  applicable to  all systems with multiorbital character of states at the Fermi energy.
\end{abstract}
\maketitle


\section{Introduction}
Discovery of the high temperature superconductors \cite{bednorz1986} a few decades ago started   
vivid and still ongoing experimental and theoretical races to uncover their secrets. This breakthrough 
in the research of superconductivity was  preceded by the discovery of CeCu$_2$Si$_2$, 
the first heavy fermion superconductor \cite{steglich1979} and other 
members of this family UBe$_{13}$ \cite{ott1983} and UPt$_3$ \cite{stewart1984}.  
The concomitant development of new technologies of material synthesis, and curiosity these findings   
have evoked, resulted in the numerous discoveries of simple, like MgB$_2$ \cite{nagamatsu2001}, and
more complex families of superconducting materials including Sr$_2$RuO$_4$ \cite{maeno1994} the perovskite superconductor
without copper. A lot of excitement have evoked dicoveries of systems superconducting  
at high pressures. These findings culminated in  recent measurements  of superconducting
materials with transition temperature $T_c$ approaching not so cold winter day temperature \cite{drozdov2015,maddury2019}.
Novel superconductors are often denoted as exotic \cite{gorkov1985,brandow1988} or (in more modern language) 
as {\it unconventional}. This last word sometimes is used in relation to superconductors  in which
other than electron - phonon pairing mechanism is operative \cite{stewart2017}. Slightly more formally,
unconventional chaeacter of superconductivity  means
that not only the gauge U(1) symmetry is spontaneously broken below the transition temperature as in all 
superconductors  but also other symmetries are broken. Among them the  time reversal symmetry (TRS) plays a special role.

Recently, the increased interest  is observed in the detection and understanding the TRS 
breaking and its relation to other symmetries. In the last few years
a number of time reversal superconductors with chiral \cite{kallin2016} and other order 
parameters breaking the time reversal symmetry have been identified.
The importance of chiral superconductors is in part related to the fact that they have been predicted 
to host Majorana particles and other quantum states of interest \cite{ivanov2001}. 
As an example of such system we consider Sr$_2$RuO$_4$ and  present some of its properties and 
our approach to model the material. We   concentrate 
on the theory of Kerr effect in this material paying special attention
to its multi-orbital character. This is because the recent discovery of the novel \cite{taylor2012,wysokinski2012} 
intrinsic mechanism of the Kerr effect operating in strontium ruthenate and possibly other systems. 
 
The paper is organised as follows. In the next section (\ref{sec:TRSBsc}) we  
review the earlier and  recent experimental discoveries of superconductors which break TRS. 
Our studies of various aspects of superconductivity in Sr$_2$RuO$_4$ are
reviewed in Section (\ref{sec:sr2ruo4}). The multiorbital/multiband mechanism of 
the Kerr effect is introduced in Section (\ref{sec:kerr}) with application to strontium ruthenate.
We end up with few general observations and the summary of the paper in Section (\ref{sec:sum}).

\section{Time reversal symmetry breaking in superconductors} \label{sec:TRSBsc}
The detection of the time reversal symmetry (TRS) breaking necessarily means the unconventional
nature of superconductivity. This is so, becouse in many  superconductors the U(1) gauge symmetry spontaneous breaking
 is often accompanied by additional symmetries: time reversal or spatial. For example, the d-wave character of 
the order parameter symmetry in high temperature superconductors (HTS), as confirmed again recently with a new 
dielectric resonator method \cite{bae2019}, provides a case of spatial symmetry breaking. The crystals of HTS 
 have a four-fold symmetry axis perpendicular to their $ab$ plane but the d-wave order parameter 
$\Delta(\mathbf{k})=\Delta_0(T)[\cos(k_xa)-\cos(k_ya)]$ features only the two-fold one.  

\subsection{Detection methods of TRS breaking}
There are a few etsblished methods to detect existence of TRS breaking (TRSB) in superconductors. These are 
muon spin rotation and relaxation ($\mu$SR), polarised neutron scattering, detection of circular currents
and polar Kerr effect \cite{schemm2017}. Recently, there appeared different additional proposals to detect the TRSB  
state experimentally. On of them is by laser pulse probe \cite{vadimov2017}, when one expects changes  
of the condensate properties around the laser heated spot. The modifications depend on the symmetry of  the
order parameter. Depending  if it is of s-wave or p-wave  one expects different patterns of the magnetic field,
which can be measured. Still another proposal relies on detection of the diamagnetic response of
multi-band superconductors \cite{yerin2017}. According to the authors the current induced magnetic flux 
response could in principle be used to to detect breaking of TRS in the ground state.
  
The method most often applied to detect breaking of TRS at the onset to the
superconducting state is  muon spin rotation and relaxation ($\mu$SR). 
If the surface of the superconductor can be suitably prepared  the polar Kerr effect can also 
be measured.   
One or both of these techniques have been applied to different superconducting systems and 
resulted in observations indicating TRS breaking in Sr$_2$RuO$_4$, UPt$_3$, URu$_2$Si$_2$
and many more superconductors as indicated in Table (\ref{tab:1}).  
Recently the epitaxial bilayers of Bi and Ni \cite{gong2017,chauhan2019} were found not only to show 
a superconductivity, but also the one with TRS breaking superconducting order parameter. Neither the mechanism nor 
the superconducting state in the system is uniquely identified. We mention, without  
deeper discussion, the  observations indicating TRS breaking near the pseudogap phase 
of HTS \cite{xia2008}, which might be caused by the recently discovered \cite{zhang2018} slowly 
fluctuating magnetic fields of intra-unit-cell origin in the same region of the phase diagram.

Superconductors which break time reversal symmetry may also have distinctive response to other 
disturbances like local temperature bias $\Delta T(r)$, etc. It has been found \cite{garaud2016} 
that the thermoelectric response of TRS breaking superconductors to $\Delta T(r)$ depends on the thermally induced magnetic 
field with a profile that is sensitive to the presence of domain walls and anisotropy of superconducting 
states. If the heating process is non-stationary the time dependent \textbf{B} field 
produces an electric field. As a result there appears a charge imbalance in different bands. In view of the puzzles related to
understanding thermopower in superconductors \cite{ginzburg1944,vanharlingen1980,barybin2008,shelly2016} and hybrid 
structures \cite{wysokinskim2012,wysokinskim2013} this is an interesting result worth of deeper analysis.

The $\mu$SR technique is a local, very sensitive probe \cite{amato1997,sonier2000} detecting tiny magnetic fields 
in the bulk of superconductors. It appears below the superconducting transition temperature. 
The beam of spin-polarised muons is directed towards the sample. The  individual, typically positive, muons injected   
into the sample  stop at the random point due to the loss of energy by electrostatic interactions. These interactions 
preserve the direction of spin polarisation of muons. After the time comparable to their lifetime ($\approx 2.2 \mu s$)
the muons in the sample decay {\it via} week interaction process into positron and two neutrinos. The positron  
is preferably emitted in the direction parallel  towards the muon spin. Thus the angular distribution 
of the positrons reflects the angular distribution of the muons. Existence of, even tiny, 
magnetic fields (of order of 0.1 G) at the stopping site of the muon induces 
the Larmor precession of its spin, what changes the measured distribution of positrons. The application of this 
technique to study unconventional superconductors have been recently reviewed \cite{bhattacharyya2018}.

To see how the Kerr effect is related to the frequency dependent Hall 
transport coefficient $\sigma_{xy}(\omega)$ let us note that   
the reflection coefficient $|r|$ 
\begin{equation}
|r|=\frac{|n-1|}{|n+1|},  \label{r}
\end{equation}
is directly related to the polar Kerr angle $\theta_K$  \cite{White-Geballe}
\begin{equation} 
\theta_K=\frac{4\pi}{\omega}\,\mathrm{Im}\frac{\sigma_{xy}(\omega)}{
n\,(n^2-1)},  \label{theta_K}
\end{equation}
where $n$ is the complex refraction coefficient. On the other hand the polar Kerr angle 
(\ref{theta_K}) can be shown \cite{lutchyn2008} to read 
\begin{equation}
\theta_K=\frac{4\pi\omega^2\,Im\sigma_{xy}(\omega)} {\sqrt{
\epsilon_\infty\omega^2-\omega_{ab}^2} \,
[(\epsilon_\infty-1)\omega^2-\omega_{ab}^2]},  \label{theta_K-high}
\end{equation}
in the high frequency regime ($\omega > \omega_{ab}$) and 
\begin{equation}
\theta_K=-\frac{4\pi\omega^2\,Re\sigma_{xy}(\omega)} {\sqrt{
\omega_{ab}^2-\epsilon_\infty\omega^2} \,
[(\epsilon_\infty-1)\omega^2-\omega_{ab}^2]},  \label{theta_K-low}
\end{equation}
for light frequencies smaller than the in-plane plasma frequency $\omega_{ab}$.
The experimental details related to the measurements of the polar Kerr effect 
in TRS breaking superconductors, together with the relevant examples of 
the temperature dependence of $\theta_K$, have been presented by Kapitulnik 
and co-workers \cite{kapitulnik2009,schemm2017}. The polar Kerr effect measures the rotation 
of the linearly polarised light reflected from the surface at normal incidence. 
The proper analysis of the Kerr effect requires careful 
consideration of the symmetries. Due to the delicate relation between 
reciprocity and time reversal symmetries a number of  proposals turned 
out to be not valid as discussed recently \cite{kapitulnik2015}. For example, 
the gyrotropy, being a result of natural optical activity of the materials can not 
lead to the non-zero Kerr response.

\subsection{Superconductors with TRS broken state}

Table (\ref{tab:1}) gives a summary of TRS breaking materials together with some of their properties.
In the table we indicate  the method(s) used to detect TRS, the superconducting transition temperature and
an important crystallographic aspect, namely the existence or not of the centre of inversion.
The  systems with the center of inversion are  known as centrosymmetric (C) and those without it 
as non-centrosymmetric (NC).
The informations on the superconducting order parameter are mostly scarce. They are subject to 
ongoing intensive research. Some of the compounds mentioned in the Table have been known for many 
years, while others have been discovered only recently. 
Here we shall add a few remarks concerning the  properties of some materials and/or their families. 
For more comprehensive discussion we direct the reader to the original cited literature.  

The superconductivity in URu$_2$Si$_2$ has been discovered \cite{schlabitz1986} in 1986, but most of the work has been devoted
to the 17.5K anomaly observed in this material and termed ``hidden'' order. The identification of
it is the subject of the on-going vigorous work and debate \cite{mydosh2011}. One of the first examples where the 
broken time reversal symmetry has been detected \cite{luke1993,brawner1997,joynt2002} is UPt$_3$.
 This superconductor has a reach phase
diagram on the temperature - magnetic field plane with three different phases (named A, B and C) 
characterised by the unique nodal structures of the superconducting order parameter \cite{adenwalla1990}. 
It is well established that the TRS is broken in one the phases.

The effect of disorder on the TRS broken state is not known in details yet. Typically the  
disorder seem to decrease the internal fields. 
In two alloy series  Pr$_{1-y}$La$_y$Os$_4$Sb$_{12}$ and Pr(Os$_{1-x}$Ru$_x$)$_4$Sb$_{12}$ the magnetic 
field appearing below $T_c$ and revealed by $\mu$SR measurements has been found to initially
decrease linearly with solute concentration \cite{shu2011}. It has been also established that Ru doping is 
considerably more efficient in decreasing magnetic field  than La doping, with a
50\% faster initial decrease. The data suggest that broken TRS state is suppressed 
for Ru concentration $x\geq 0.6$ but persists for essentially all La concentrations. 
The detailed changes of the superconducting properties of the alloys are needed to understand this behaviour.

\begin{table}
\caption{Time reversal symmetry breaking superconductors and their properties. We indicate the method 
of detection of TRS breaking, the structure of the material if it is centrosymmetric (C) or non-centrosymmetric (NC), 
the $T_c$. The last entry provides the remarks concerning the structure
of the order parameter. 
}
\centering
\begin{tabular}{|l|c|c|c|c|}
\textbf{material}&\textbf{detection}	& \textbf{structure}	& \textbf{$T_c$}[K]& \textbf{ sc state}\\
Sr$_2$RuO$_4$	  &$\mu$SR \cite{luke1998}, Kerr \cite{xia2006}& C			& 1.5& nodal, spin triplet, $p_x\pm ip_y$ \cite{maeno1994} \\
UPt$_3$	        & $\mu$SR \cite{luke1993}, Kerr \cite{schemm2014} & C			& 0.54& spin triplet \\
URu$_2$Si$_2$   &  Kerr \cite{schemm2015}  & C			& 1.5 &  \cite{schlabitz1986}\\
PrOs$_4$Sb$_{12}$&$\mu$SR \cite{aoki2003}, Kerr \cite{levenson2018}& C& 1.8& \cite{bauer2002} \\
PrPt$_4$Ge$_{12}$&$\mu$SR \cite{maisuradze2010}& C& 7.9& \\
Pr$_{1-y}$La$_y$Os$_4$Sb$_{12}$&$\mu$SR \cite{shu2011}& C &  $\sim$ 1&  \\
Pr(Os$_{1-x}$Ru$_x$)$_4$Sb$_{12}$ & $\mu$SR \cite{shu2011}& C & &  \\ 
Ba0.27K0.73Fe2As2 & $\mu$SR \cite{grinenko2017} & C & 13 &$s+is$ or $s+id$ (?)\\
LaNiGa$_2$	    & $\mu$SR \cite{hillier2012}  & C			  & 2.1 & \\
Re& $\mu$SR \cite{shang2018}&C&2.7& fully gapped\\
Re$_{0.82}$Nb$_{0.12}$& $\mu$SR \cite{shang2018}&C&8.8& \\
SrPtAs		&$\mu$SR \cite{biswas2013}  & NC (locally)			& 2.4 &$d+id$\\
LaNiC$_2$	      & $\mu$SR \cite{hillier2009,quintanilla2010}  & NC			& 2.7 & multigapped, non-unitary triplet\\
Lu$_5$Rh$_6$Sn$_{18}$& $\mu$SR \cite{bhattacharyya2015}& C&4.0& $d+id$(?)\\
La$_7$Ir$_3$    &$\mu$SR \cite{barker2015}    & NC			& 2.25 &\\
La$_7$Rh$_3$		& $\mu$SR \cite{singh2018a}      & NC			& 2.65 &\\
Re$_6$Zr				&$\mu$SR \cite{singh2014,pang2018}    & NC			& 6.75 &\\
Re$_6$Hf				&$\mu$SR \cite{singh2017}    & NC			& 5.91 & s-wave\\
Re$_6$Ti				&$\mu$SR \cite{singh2018}    & NC			& 6.0 &\\
Re$_{24}$Ti$_5$& $\mu$SR \cite{shang2018}      & NC		& 6   & s-wave\\
Zr$_3$Ir				&$\mu$SR \cite{shang2019a}  & NC	    & 2.2  & single gap, nodeless\\
Pr$_{1-x}$Ce$_x$Pt$_4$Ge$_{12}$& $\mu$SR  \cite{zhang2015}& NC?& & \\ 
Bi/Ni (bilayers)  & Kerr \cite{gong2017}   & NC   & 3.6 & $d_{xy}\pm id_{x^2-y^2}$ (?)\\
\end{tabular}
\label{tab:1}
\end{table}
It is visible from Table (\ref{tab:1}) that many  superconductors  
with Re as one of the components belong to NC compounds and at the same time break TRS. This raises 
the question about the possible relation between both symmetries and the reasons while 
so many Re$TM$ ($TM$- transition metal) alloys feature such behaviour.  
To elucidate the issue, the authors \cite{shang2018} have performed the comparative studies of
elemental  Re  superconductor ($T_c=2.7K$) with the center of inversion and the 
Re$_{0.82}$Nb$_{0.12}$ alloy ($T_c=8.8K$) without it. 
Both superconductors have been studied with the $\mu$SR technique and small but well defined magnetic fields 
were detected below $T_c$. Such behaviour point out into the spontaneous time reversal symmetry breaking 
in both superconductors. The conclusion of the paper is that the lack of the inversion symmetry is not essential 
for the appearance of TRS breaking. This is also supported by the data presented in the Table, where many  of the 
TRS breaking materials belong to C class. On the other hand the lack of TRS strongly constraints the allowed 
symmetry of the superconducting order parameter. This together with the temperature dependence of the specific 
heat and other thermodynamic and transport characteristics allows to judge the presence of the gaps in 
the order parameter. The details of the symmetry of the order parameter in elemental Re superconductor and 
its superconducting compounds are still unknown. However, the lack of TRS in Re compounds seem to be crucially 
related to the presence of this element.   
To further explore the role of Re in the Re$TM$ compounds the binary Re$_{1-x}$Mo$_x$ alloy series 
have been prepared and shown \cite{shang2019} to be superconducting for all concentrations $x$ of Mo 
with the highest $T_c=12.4$K for $x\approx 0.6$. Depending on $x$ different crystallographic structures 
have been found. Among them the non-centrosymmetric  $\alpha$-Mn has been obtained.
 Superconductivity in non - centrosymmetric materials has been  reviewed recently \cite{smidman2017}.

\section{Sr$_2$RuO$_4$: puzzles, solutions and still open issues} \label{sec:sr2ruo4}

The superconductivity in strontium ruthenate has been discovered 25 years ago 
and its understanding still presents a challenge. The title of the recent 
review ``{\it Even odder after twenty-three years: the superconducting order parameter puzzle
of Sr$_2$RuO$_4$}'' very well summarizes \cite{mackenzie2017} the state of the art in our understanding of its  
superconducting properties. The normal state seem to be of the Fermi liquid variety and well behaved,
although the issues related to the strength of electron correlations and spin-orbit coupling are 
not quite clear \cite{pchelkina2007,rozbicki2011,haverkort2008,facio2018}. Beside the above mentioned, 
 many excellent reviews on all aspects related to strontium ruthenate exist 
\cite{kallin2016,mackenzie2003,kallin2012,maeno2012,lichtenberg2002}. Thus we shall concentrate on a few less 
common issues, namely observation and understanding of the Kerr effect in the material, recent 
experimental signatures of the multi-band and multi-condensate features.

After the discovery of the superconductivity in Sr$_2$RuO$_4$, the  
material being a crystallographically identical to high temperature 
cuprate superconductors but without copper, the hope was that its understanding will shed
light on the latter systems. This was supported by the observation that 
at low temperature strontium ruthenate is a metal and behaves as an anisotropic and possibly correlated
but otherwise well defined Fermi liquid \cite{hussey1998}. 
Later on it turned out that Sr$_2$RuO$_4$ is much more complicated with three bands being in play 
hosting probably spin triplet p-wave superconductivity. The first hint on the 
unconventional character of superconductivity has been provided by the extreme 
sensitivity of $T_c$ on non-magnetic disorder \cite{mackenzie1998}.

The Fermi surface of the materials consists of three sheets known 
as  $\alpha$, $\beta$ and $\gamma$. The first two are of one dimensional origin and result 
from hybridised $d_{xz}$ and $d_{yz}$ Ru orbitals, while the last one from $d_{xy}$ orbitals. 
This variety of orbitals has resulted in a discussion about the active bands and the role 
various orbitals play in the superconducting state \cite{agterberg1997,zhitomirsky2001,konik2007}. 
To understand experimental results showing the existence of line nodes, or at least deep minima in the quasiparticle
density of states  it has been proposed \cite{zhitomirsky2001} that a full gap exists in the active band, which 
is that derived from $d_{xy}$ orbitals and the line nodes develop in passive bands by the interband proximity effect.
The multiband aspects of superconductivity in this system, however, are of different variety 
from that considered in other materials \cite{zegrodnik2014,wysokinski2016,dai2019}, albeit 
some similarities can be found. The discussion about the role of active {\it vs.} passive bands 
continues \cite{raghu2010,wang2013,gao2013}. It is not unrelated to the studies of spin fluctuations in the material 
and the relative role of Coulomb interactions in different orbitals, mentioned earlier.

Kallin and Berlinsky finishing their review on chiral superconductors \cite{kallin2016} 
write {\it one would also like to have detailed information of where the low-lying excitations in  
Sr$_2$RuO$_4$ are in momentum space [...]}. Probably the recent experiments \cite{akebi2019} 
provide, at least partial, answer to that issue. Thus, even though the origin of spin triplet pairing in strontium ruthenate 
is still under debate, the discovery of the low energy modes \cite{akebi2019} seem to give novel argument in favour of  
the early proposals \cite{rice1995,mazin1997} pointing out at the ferromagnetic spin fluctuations operating in the material 
and being responsible for its superconducting instability. Spin fluctuations have been studied 
in numerous papers \cite{sidis1999,braden2002,braden2004,iida2011} without clear cut conclusions.
 The relative importance of the ferro- {\it vs.} antiferromagnetic
fluctuation has also been probed \cite{ortmann2013}. The conclusion of the recent paper \cite{steffens2019}
 studying this issue by means of polarized inelastic neutron scattering is that 
{\it spin fluctuations alone are not enough to generate a triplet state}. This is  in opposition to  
the results presented by Akebi {\it et al.}\cite{akebi2019} and calls for further  detailed analysis.

\subsection{Modelling of strontium ruthenate}

In view of the controversies about the microscopic mechanism of superconductivity in 
strontium ruthenae we have modelled the system by using precise knowledge related 
to its spectrum and assumed the phenomenological interaction parameters leading to 
the p-wave superconducting state in all three bands. The details have been described in the
number of papers \cite{Annett2002,Annett2003,wysokinski2003,Annett2006,Annett2009,wysokinski2009} 
and I shall not repeat the details here.
The Hamiltonian of the system written in the orbital basis reads
 \begin{eqnarray}
 \hat{H}& =&  \sum_{ijmm^\prime,\sigma}  \left( ( \varepsilon_m -
 \mu) \delta_{ij} \delta_{mm^{ \prime}} - t_{mm^{ \prime}}(ij)  \right)  \hat{c}%
^+_{im \sigma} \hat{c}_{jm^{ \prime} \sigma}   \nonumber  \\
 & & + \mathrm{i}  \frac{ \lambda}{2}  \sum_{i, \sigma \sigma^{ \prime}}
 \sum_{mm^{ \prime}}  \varepsilon^{ \kappa mm^{ \prime}}
 \sigma^{ \kappa}_{ \sigma \sigma^{ \prime}} c^+_{i m \sigma} c_{i
m^{ \prime} \sigma^{ \prime}} \nonumber \\
&& -  \frac{1}{2}  \sum_{ijmm^{ \prime}}
U_{mm^{ \prime}}^{ \alpha \beta, \gamma \delta}(ij) 
 \hat{c}^+_{im \alpha } \hat{c}^+_{jm^ \prime \beta } \hat{c}_{jm^ \prime \gamma } 
 \hat{c}_{im \delta },
 \label{hubbard}
 \end{eqnarray}
 where $m$ and $m^{ \prime}$ refer to the three Ru $t_{2g}$ orbitals 
$a=d_{xz}$, $b = d_{yz}$ and $c = d_{xy}$; $i$ and $j$ label the sites of a body
centered tetragonal lattice. The hopping integrals $t_{mm^{ \prime}}(ij)$ and
site energies $ \varepsilon_m$ were fitted to reproduce the experimentally
determined Fermi surface  \cite{mackenzie1996,bergemann2000}. $\lambda$ is the effective 
Ru 4d spin-orbit coupling parameter, and the effective Hubbard  parameters 
$U_{mm^{ \prime}}^{ \alpha \beta, \gamma \delta}(  ij)$  are generally spin 
as well as orbital dependent. In its simplest version the  model uses just two 
effective Hubbard interaction parameters,  both of which are completely 
determined by the requirement to fit the experimental $T_c$. One of the
parameters is responsible for the effective attraction of electrons in $d_{xy}$ orbitals
of in plane Ru atoms, while the other is for attraction between electrons in out of 
plane Ru orbitals.

One of the puzzling behaviours of strontium ruthenate has been found in measurements of the electronic spin susceptibility 
 in  external magnetic field parallel and perpendicular to the {\it ab} plane of the material. The spin susceptibility
of spin-triplet superconductors is known to have matrix (3$\times$3) structure with entries depending on the direction of 
the \textbf{d} - vector. For standard spin-triplet odd parity chiral state the d-vector is of the 
form \cite{leggett1975,annett1990}
\begin{equation}
              {\bf d}({\bf k}) = (\sin{k_x}+i\sin{k_y})\hat{\bf e}_z,
          \label{eq:chiral-a}
\end{equation} 
and is directed along crystallographic c-axis, while for the helical states of the variety 
\begin{eqnarray}
  {\bf d}({\bf k}) &=& (\sin{k_x},\sin{k_y},0)    \\
   {\bf d}({\bf k}) &=& (\sin{k_y},-\sin{k_x},0)  \nonumber
\end{eqnarray} 
it is lying in the {\it (a,b)}-plane of the strontium ruthenate. 
The early $^{17}O$ Knight shift experiments performed in (ab) plane magnetic fields \cite{ishida1998}
and neutron scattering  experiments \cite{duffy2000} both observed
a constant susceptibility below $T_c$  consistent with chiral triplet paring state with 
${\bf d}$ along c-axis. On the other hand the measurements by  
Murakawa {\it et al.} \cite{murakawa2004}   in magnetic field  parallel to the c-axis also observed
spin susceptibility below $T_c$ of the same value as in the normal state. This immediately indicates 
that either the superconducting state is not chiral, or \textbf{B} field induces the  phase transition from the chiral
to helical  state. Using the three band model with relatively small but realistic spin-orbit coupling we
have argued \cite{annett2008,annett2007} that the d-vector rotates and the phase transition is expected. The calculated
entropy jump at low temperature has been found to be very small, so the transition could be undetected in specific
heat experiments.

\subsection{Horizontal or vertical line nodes?}

Small angle neutron scattering studies \cite{kuhn2017} provide support to the anisotropic multiband 
superconducting state with gap nodes or at least deep minima. The authors inferred the multiband behaviour 
from   the superconducting anisotropy in Sr$_2$RuO$_4$ which is hardly temperature dependent  but increases 
for higher  fields ($\geq$ 1 T). This paper found evidence from vortex lattice distortion 
that the intrinsic superconducting anisotropy between the c axis and the Ru-O basal 
plane being of the order $60$ is in agreement with that measured \cite{kittaka2014} by the ratio 
of coherence length $\xi_{ab}/\xi_c\approx 60$, but exceeds 
the magnetic field anisotropy  $H_{c2}^{ab}/H_{c2}^c\approx 20$.
In line with the above findings is the work of Kallin and co-workers who have shown \cite{dodaro2018} 
that if there are no horizontal nodes in the superconducting 
order parameter of Sr$_2$RuO$_4$ so it is particularly difficult to reconcile chiral-p-wave order 
with residual thermal conductivity data. The model (\ref{hubbard}) leads to the horizontal nodes 
in the superconducting order parameter. The order parameter   has a number of intra- and interorbital components.
They are written as 
\begin{equation}
\Delta _{cc}({\bf k})=\Delta^x_{cc}(T)\sin {k_{x}}\pm i\Delta^y_{cc}(T)\sin{k_{y}})
\label{eq.delta2}
\end{equation}
for $c (=d_{xy})$ orbitals and,
\begin{equation}
\Delta _{mm^{\prime}}({\bf k})= 
(\Delta^x_{mm^{\prime}} 
\sin {\frac{k_{x}}{2}}\cos {\frac{k_{y}}{2}}\pm i \Delta^y_{mm^{\prime}}\sin {\frac{k_{y}}{2}}\cos {\frac{k_{x}}{2}}
)\cos {\frac{k_{z}c}{2}}  
\label{eq.delta3}
\end{equation}%
for $m,m^{\prime }=a,b$ or $d_{xz}$ and $d_{yz}$ orbitals forming $\alpha$ and $\beta$ Fermi sheets. 
The experiments \cite{hassinger2017} showing the existence of vertical line nodes 
in Sr$_2$RuO$_4$ provide a good reason for further analysis of the gap nodes  as 
it is not clear if the vertical anisotropy of the gaps  as given by (\ref{eq.delta3}), together 
with the warping of the essentially cylindrical  Fermi surface  is enough
to understand the measurements. For  recent group - theoretical analysis
of the gap nodes in tetragonal crystal see \cite{yarzhemsky2018}.

\subsection{Surface magnetic fields}
As already indicated, the  $\mu$SR and the polar Kerr effect  experiments point towards appearance
of the spontaneous magnetic fields inside the chiral strontium ruthenate superconductor.
According to the expectations  for such superconductors at their surfaces, at domain walls and 
near the impurities \cite{sigrist1991} one should observe small magnetic field. 
Despite many experimental efforts none of the measurements
 has found such surface  magnetic fields \cite{kirtley2007,hicks2010,curran2014}. For example in the 
paper \cite{kirtley2007} the authors  used sensitive scanning Hall bar and superconducting quantum 
interference device microscopies  and did not detect expected supercurrents.  
Negative results have been obtained in similar measurements for Sr$_2$RuO$_4$4 and 
PrOs$_4$Sb$_{12}$ \cite{hicks2010,aoki2007} and suggested that the size of the chiral domains might
be much smaller than expected. The paper \cite{curran2014}  imposes an upper limit of 
$\pm$2.5 mG on the magnitude of spontaneous magnetic fields at the well-defined edges 
of a mesoscopic disk. It is important to note that despite lack of surface fields the superconductivity-related time-reversal 
symmetry-breaking fields in the bulk have been observed by muon spin rotation and Kerr 
effect (see Table (\ref{tab:1})). Judging from the value of the Kerr angle the fields seem to be really small.

On the theory side there appeared a number of papers trying to find reasons 
of such behaviour \cite{ashby2009,lederer2014,huang2015}. Recent studies \cite{etter2018} have 
shown that the surface flux pattern in chiral 
superconductors is not a universal feature, but instead it depends on many 
parameters describing the system. As a consequence the magnitude of the expected 
magnetic fields may differ from case to case and be smaller than expected earlier \cite{matsumoto1999}.

\section{Understanding the Kerr effect in Sr$_2$RuO$_4$ - multiorbital mechanism}\label{sec:kerr}

Shortly after the measurements of the Kerr signal in Sr$_2$RuO$_4$ a number of theoretical 
papers appeared  trying to understand its origin, magnitude and temperature dependence. 
For  references and a critical discussion see \cite{lutchyn2008,lutchyn2009}. As noted earlier 
the existence of the Kerr signal is intimately related to the  anomalous frequency dependent 
Hall effect $\sigma_H(\omega)$. Due to the time reversal symmetry breaking state in 
superconductors one expects appearance of spontaneous magnetic fields and thus the Hall effect. The latter 
transport coefficient, however, requires that the charge in an electric field directed along, 
say $x$ direction, to move also in $y$ direction  
as in standard Hall effect in a \textbf{B} field. In the presence of magnetic field the perpendicular ($i.e.$ in the $y$ diection) 
motion is related to the Lorentz force acting on However, in the present case the chiral state responsible for time reversal 
breaking state is a result of internal interactions between the electrons. 
Thus the required force has to result from effective internal interactions inducing the chiral $k_x\pm i k_y$ state.
This, however is impossible in the Galilean invariant system. This argument about the absence of such skew scattering 
in one - band  Galilean invariant system has been put forward by Read and Green \cite{read2000}. Good discussion of 
different constraints related to the observation of Kerr effect can  be found in \cite{kapitulnik2015}. Later on on we 
shall present more formal argument showing vanishing of the Hall conductivity in one band chiral superconductor.

It is important to remind that the breaking of time reversal symmetry is a necessary 
but not sufficient condition for the   observation of the Kerr effect. This is also true 
for magnetic systems, where spin-orbit interaction is a required additional ingredient \cite{ebert1996} for
the Hall effect to exist. It turns out 
that similar situation is observed in the time reversal symmetry breaking superconductors. 
To observe the Kerr effect additional factors have to contribute.  In the literature
there were various mechanism discussed. These included particle - hole asymmetry \cite{yip1992}, the 
 order parameter collective mode response \cite{roy2008}, the impurity scattering leading to 
non-trivial vertex corrections and multiorbtal effects and the final size of the laser spot.

The ac Hall effect in chiral superconductors require breaking of the translational invariance 
and this is achieved by impurities as proposed originally by Goryo \cite{goryo2008} and later elaborated
by Lutchyn and coworkers \cite{lutchyn2008,lutchyn2009}. 
Other possibility is provided by the many orbital or many band models. Such novel, earlier
not considered mechanism of the Kerr effect has been independently proposed by two 
groups \cite{taylor2012,wysokinski2012}. 

The simple, formal argument on the absence of ac Hall conductivity $\sigma_H(\omega)$ 
has been given by Taylor and Kallin \cite{taylor2012}.  These authors have noted that 
$\sigma_H(\omega)$  is given by the antisymmetric part 
of the current - current correlation function $\pi_{xy}({\bf q},\omega)$
\begin{equation}
\sigma_H(\omega)=\frac{i}{2\omega}\lim_{q\rightarrow 0}[\pi_{xy}({\bf q},\omega)-
\pi_{yx}({\bf q},\omega)].
\label{hall-kubo}
\end{equation}
The correlator itself can be written in terms of the velocity matrices
$\hat{\mathbf v}_x$ and $\hat{\mathbf v}_x$ 
and matrices of the Green functions $\hat{\mathbf G}$ of the superconducting system as
\begin{equation}
\pi_{xy}({\bf q},i\omega_n)=e^2T\sum_{{\bf k},i\omega_l}tr\left[
\hat{\mathbf v}_x\hat{\mathbf{G}}({\bf k},i\omega_l)\hat{\mathbf v}_y\hat{\mathbf{G}}({\bf k}+{\bf q},i\omega_l+i\omega_n)\right].
\label{diag-vel}
\end{equation}
It follows that $\sigma_H$ vanishes for diagonal  velocity matrices $\hat{\mathbf{v}}_x$ and $\hat{\mathbf{v}}_y$ 
as they commute with  the  Green function matrices. 
The critical overview of various attempts to calculate Kerr effect can be found in \cite{mineev2010}.

Two different approaches have been used and two different models
sharing, however, the multiband/multiorbital character have been studied by the two 
groups \cite{taylor2012,wysokinski2012} proposing the novel mechanism of the Kerr effect. 
The results \cite{taylor2013,gradhand2013,gradhand2015} 
provide a novel description of the anomalous ac Hall conductivity and are valid for 
virtually all superconductors. The prerequisite is the TRS breaking order parameter
and a multiband one with non-zero interorbital/interband velocity matrices.  

While the Kubo approach has been used in \cite{taylor2012}, we  \cite{wysokinski2012} have 
calculated the Kerr effect from the definition of the optical dichroism  \cite{capelle1997,capelle1998}.
 In this formalism the conductivity tensor can be expressed in terms of the difference of the 
electromagnetic power absorption ${P}(\omega ,\vec{\epsilon})$ for light of
left and right circular polarizations, $\vec{\epsilon}_{L}$ and $\vec{\epsilon}_{R}$,
 respectively, 
\begin{equation}
\mathrm{Im}[\sigma _{xy}(\omega )]=\frac{1}{VE_{0}^{2}}\left[ {P}(\omega ,
\vec{\epsilon}_{L})-{P}(\omega ,\vec{\epsilon}_{R})\right] .  \label{imsxy}
\end{equation}
The power has been calculated from the Fermi golden rule with the dipole 
matrix elements evaluated between the Bogolubov-de Gennes states.

The interorbital mechanism seem to be especially well suited
for the understanding of the Kerr effect in strontium ruthenate, which is perhaps the
cleanest superconductor, ever studied and the concurrent mechanism relying on higher order impurity
scattering is not efficient. The novel mechanism of polar Kerr effect discovered during 
studies of the Kerr effect in this superconductor is of general importance and   its validity 
for the semi-quantitative description of the effect in UBe$_3$ has been recently demonstrated \cite{wang2017}.

\section{Summary}\label{sec:sum} 
We have reviewed aspects related to time reversal symmetry breaking in superconductors 
including Sr$_2$RuO$_4$, UPt$_3$ and other newly discovered systems.
One has to note that the family of materials with this property is  growing very fast.
Moreover, many compounds belong to the non-centro-symmetric crystals. This calls for better
understanding of the interplay between various symmetries in superconductors.

We briefly described  standard techniques 
used to identify TRS breaking in superconductors, namely  the $\mu$SR and the Kerr effect.
 We have discussed  some of the many puzzling characteristics of
strontium ruthenate (Sr$_2$RuO$_4$), one of the cleanest and best studied superconductors with TRS breaking state 
below $T_c$. The special attention has been paid to the discovery of the novel mechanism 
of the Kerr effect \cite{taylor2012,wysokinski2012} and its application to Sr$_2$RuO$_4$.
However, many of  the recent discoveries  
\cite{wu2016,koenig2017,robbins2017,anwar2017,watson2018,yang2018,huang2018,wang2018,zhang2018a,wu2019,komendova2017,triola2018} 
have been left, partly due to lack of space in this short review. We only mention that the recent work \cite{komendova2017} 
has established close relation between the Kerr rotation and odd-frequency superconductivity. Both are emerging from the same 
finite hybridization between different orbitals. Thus Sr$_2$RuO$_4$ appears as one of the first bulk materials hosting 
odd-frequency superconductivity.

The discussed multiorbital mechanism of the Kerr effect is universally valid in both clean and 
dirty  materials. In view of the recent interests in the study of TRS breaking  in 
superconductors it would be of great interest to evaluate the relative share  
of the impurity and inter-orbital contributions to the measured signal. This should be
possible by means of the controlled disordering and the concomitant measurements of the 
Kerr angle  of one of the many novel superconductors  breaking the TRS.

\acknowledgments{The work reported here has been supported by the 
M. Curie - Sk\l{}odowska University and the National Science Centre 
(NCN) grant DEC-2017/27/B/ST3/01911 (Poland). I would like to thank
J. F. Annett, G. Litak and M. Gradhand for collaboration on some problems
tackled in this paper.}


\end{document}